\documentclass[preprint]{aastex}

\slugcomment{to appear in ApJ Letters}
\begin{document}
\title{A Search for Distant Solar System Bodies in the Region of Sedna}
\author{Megan E. Schwamb \altaffilmark{1},Michael E. Brown\altaffilmark{1}, and David L. Rabinowitz \altaffilmark{2}}
\altaffiltext{1}{Division of Geological and Planetary Sciences, California Institute
of Technology, Pasadena, CA 91125}
\altaffiltext{2}{Department of Physics, Yale University, P.O. Box 208121, New Haven, CT 06520}
\email{mschwamb@gps.caltech.edu}

\begin{abstract}
We present the results of a wide-field survey for distant Sedna-like bodies in the outer solar system using the 1.2-m Samuel Oschin Telescope at Palomar Observatory. We searched $\sim$12,000 square degrees down to a mean limiting magnitude of 21.3 in R. A total number of 53 Kuiper belt objects and Centaurs have been detected; 25 of which were discovered in this survey. No additional Sedna-like bodies  with perihelia beyond 70 AU were found despite a sensitivity to motions out to $\sim$1000 AU. We place constraints on the size and distribution of objects on Sedna orbits.
\end{abstract}
\keywords {Kuiper Belt - Oort Cloud - solar system: formation}
\section{Introduction}
The discovery of Sedna \citep{2004ApJ...617..645B} suggests the presence of  a population of icy bodies residing far outside the Kuiper belt. Sedna is dynamically distinct from the rest of the Kuiper Belt. With a perihelion of 76 AU, Sedna is well beyond the reach of the gas-giants and, unlike typical Kuiper belt objects (KBOs), could not be scattered into its highly eccentric orbit from interactions with Neptune alone \citep{2003MNRAS.338..443E, 2005CeMDA..91..109G}. The orbits of many scattered KBOs extend well beyond SednaÕs perihelion, but their perihelia remain coupled to Neptune below 50 AU. Sedna's aphelion at $\sim$1000 AU is too far from the edge of the solar system to feel the perturbing effects of passing stars or galactic tides in the present-day solar neighborhood \citep{1987AJ.....94.1330D,1997Icar..129..106F}.  Some other mechanism no longer active in the solar system today is required to emplace Sedna on its orbit.

Several formation mechanisms have been proposed to explain Sedna's origin, including interactions with planet-sized bodies \citep{2008AJ....135.1161L, 2006ApJ...643L.135G, 2006Icar..184..589G}, stellar encounters \citep{2004AJ....128.2564M}, multiple stellar fly-bys in a stellar birth cluster \citep{2006Icar..184...59B, 2007Icar..191..413B,2008Icar..197..221K}, interstellar capture \citep{2004Natur.432..598K,2004AJ....128.2564M}, and perturbations from a wide-binary solar companion \citep{2005EM&P...97..459M}. The study of the Sedna population provides a unique new window into the history of the early solar system.  Each of the proposed scenarios leaves a distinctive imprint on the members of this class of distant objects and has profound consequences for our understanding of the solar system's origin and evolution. The orbits of these distant planetoids are likely dynamically frozen in place providing a fossilized record of their formation. Sedna is the only body known to reside in this region. To date, wide-field surveys \citep{2008ssbn.book..335B, 2007AJ....133.1247L}, have been unsuccessful in finding additional Sedna-like bodies. 

\section{Observations}
From the wide-field survey in which Sedna was discovered, \citet*{2008ssbn.book..335B} estimated that between 40-120 Sedna-sized bodies may exist on similar Sedna-like orbits. In order to find additional members of this population, we have been engaged in an observational campaign to survey the northern sky for both fainter and more distant objects. From 2007 May 8 - 2008 September 27, we have surveyed 11,786 square degrees within $\pm$ 30 degrees of the ecliptic (see Figure 1) to a mean depth of R magnitude 21.3. In this paper, we present the preliminary results of our survey and place constraints on the size of a distant Sedna population. 

Observations were taken nightly using the robotic 1.2-m Samuel Oschin Telescope located at Palomar Observatory and the QUEST large-area CCD camera \citep{2007PASP..119.1278B}. The QUEST camera has an effective field of view of 8.3 square degrees with a pixel scale of 0.87$^{\prime\prime}$. The 161-megapixel camera is arranged in four columns or ``fingers'' along the East-West direction each equipped with 28 2Kx640 CCDs in the North-South direction. The gap between chips in the North-South direction is $\sim$1.2$^{\prime}$, and the spacing between adjacent fingers along the East-West direction is  $\sim$35$^{\prime}$. The RA chip gap is covered by adjacent pointings, but the declination gap remains mostly uncovered. 

We observe over a two-night baseline to distinguish the extremely slow motions of distant Sednas from background stars. For each target field, a pair of 240s exposures is taken separated by  $\sim$1 hour on each of the two nights.  The second night of observations is typically the next day or at most four nights later. All exposures are taken through the broadband RG610 filter. Target fields are observed within 42 degrees of opposition where the apparent movement of these objects is dominated by the Earth's parallax. If all opposition fields for a month's lunation have been completed, overlap pointings are then targeted to reduce holes in our sky coverage due to the camera's declination gap and defective CCDs. 

Images are bias subtracted and flat field corrected.  A flat field for each of the CCDs is constructed from a median of the night's science frames. Each CCD is searched separately for moving objects. Sextractor \citep{1996A&AS..117..393B} is used to generate a list of all sources in each image. Sources that have not moved within a 4-arcsecond radius between the two nights are removed as stationary background stars. Potential moving candidates are then identified from the remaining unmatched sources. The nightly images are searched for moving object pairs with motions less than 14.4 $^{\prime\prime}$hr$^{-1}$, the velocity of bodies at distances of 10 AU or greater.  Moving object pairs from  the first night and pairs from the second night with consistent magnitudes and velocities are linked. Candidates with apparent prograde motion between the two nights, inconsistent with opposition, are rejected.  Distant objects may move too slowly to show apparent motion over one-hour baselines. We allow candidate objects to appear stationary on individual nights; we only require motion to be identified over the two-night baseline. 
 
 To further reduce the number of false positives, candidates are filtered via the orbit-fitting package described in \citet*{2000AJ....120.3323B}. Those candidates with best fit orbits producing a chi-squared less than 25 and a barycentric distance greater that 15 AU are screened by eye. To confirm there is a moving source present, the discovery images of these candidates are aligned and blinked. Recovery observations are performed within the first three months of discovery on all final moving object candidates to remove contamination from slow-moving asteroids near their stationary points and faint background stars at the limiting magnitude. One-year recovery observations are still ongoing for our new discoveries. 

Observations are taken during a wide variety of photometric, seeing, and weather conditions. Each CCD frame is calibrated independently. For each image we derive a least squares best-fit magnitude zero offset to our instrumental magnitudes relative to the USNO A2.0 catalog \citep{1998usno.book.....M} red magnitude. The photometric uncertainty of the USNO catalog is non-negligible. For magnitudes greater than 16, the uncertainty is 0.3 mag \citep{1998usno.book.....M}. We likely have several tenths of magnitude uncertainty in our discovery magnitudes. We have not fully calibrated the survey depth, but the average limiting magnitude based on the USNO catalog is 21.3 in R. 

The original Palomar survey \citep{2003EM&P...92...99T, 2008ssbn.book..335B}, which discovered Eris and Sedna, was sensitive to motions out to 1$^{\prime\prime}$hr$^{-1}$  ($\sim$150 AU) and a limiting magnitude of $\sim$20.5 in R. Our survey can detect motion out to $\sim$1000 AU ($\sim.2^{\prime\prime}$hr$^{-1}$) and probes almost a full magnitude deeper than the previous Palomar survey. We are sensitive to Mars-sized bodies out to a distance of $\sim$300AU and to Jupiter-sized objects residing at $\sim$1000 AU. 

\section{Results and Analysis}

A total number of 53 KBOs and Centaurs have been detected, of which 25 are new discoveries from this survey. The radial distribution of our detections is plotted in Figure 2. Of the objects found in our survey only two reside past 80 AU: Sedna and 2007 OR10 (discovered in this survey). All known objects past 80 AU within our magnitude limit were detected except for Eris. On both nights, Eris was located in the $\sim$1.2 arcminute declination gap between the CCDs and therefore was not positioned on any of our images. 2007 OR10 was detected moving at 1.4$^{\prime\prime}$hr$^{-1}$ at a barycentric distance of 85.369 $\pm$ 0.004. With an R magnitude of 21.4, this object is almost a full magnitude fainter than Sedna (R=20.7). 

From the discovery observations alone 2007 OR10 cannot be identified as a Sedna-like body on a high-perihelion orbit. Many scattered KBOs have aphelia well outside the planetary region past 50 AU. Both families of orbits provide reasonable fits to the short discovery arc. The two orbital solutions diverge sufficiently within a year after discovery, and a secure dynamical identification can only be made after these additional observations. Follow-up observations from the Palomar 60-inch telescope and the 0.9-m SMARTS telescope at Cerro Tololo between July 2007 and August 2008 confirm that 2007 OR10 is a scattered disk KBO close to aphelia. The best fit orbit yields a semi-major axis of a=66.99 $\pm$ 0.06 AU, an eccentricity of e=.503 $\pm$ 0.001, and an inclination of i=30.804 $\pm$ 0.001 degrees. 

No new Sedna-like bodies with perihelia beyond 70 AU were found in the survey despite a sensitivity out to distances of $\sim$1000 AU. An object of Sedna's size and albedo would have been detected up to a distance of $\sim$93 AU. To place constraints on the number of bodies in the Sedna region, we developed a survey simulator to compare the expected number of detections from a theoretical population to our survey results. The simulator draws synthetic objects from a model orbital and absolute magnitude  distribution and for every image computes the positions  and brightnesses of these objects on the sky. Those synthetic objects that lie within our sky coverage with an apparent magnitude above both nights' limiting magnitudes are considered valid survey detections. Objects have multiple detection opportunities due to repeat sky coverage over subsequent years and overlapping fields. We do not count duplicate detections in our tallies. 

We model a population of bodies on Sedna orbits with the same semi-major axis and eccentricity as Sedna (a= 495 AU  e=.846 ) and randomize over all other orbital angles. Inclinations are selected from an inclination distribution adapted from  \citet*{2001AJ....121.2804B} having a functional form of:
\begin{equation}
{\rm N}(i \le i_{max})=\int_0^{i_{max}}  { \rm exp} \left({\frac{-i^2}{2\sigma^2}}\right)\sin({i)} \,{\rm d}i
  \end{equation} 
 where $\sigma$ is chosen to be 10.25 to make Sedna's inclination of 11.9 degrees the median inclination. Two million objects are drawn from our theoretical Sedna population. Approximately half of the synthetic Sednas are located within our sky coverage. 

Due to the large uncertainties in the albedo distribution of such a distant population, we assign absolute magnitudes to our synthetic bodies instead of diameters. We assume a single power-law brightness distribution similar to the Kuiper belt where the number of objects brighter than a given absolute magnitude, H$_{max}$, is described by:
\begin{equation}
 {\rm N} (H \le H_{max)}=N_{H\leq1.6}10^{ \alpha (H_{max}-1.6)}
 \end{equation}
   The brightness distribution is scaled to $N_{H\leq1.6}$, the number of bodies with an absolute magnitude brighter than or equal to Sedna (H=1.6). For these simulations, we use a value of $\alpha=. 58$ as measured for a single-power law fit to the Kuiper belt by \citet*{2009AJ....137...72F}. 
      
For each value of   $N_{H\leq1.6}$ between 1 and 250, we perform 100,000 survey simulations and tally the number of simulations in which, like the real survey, one object on a Sedna-like orbit is detected.  Absolute magnitudes are randomly assigned to our simulated Sednas 100,000 times, for every value of  $N_{H\leq1.6}$, A single instance of the brightness distribution can be thought of as a separate survey. For each  $N_{H\leq1.6}$ tested, the number of  ``surveys'' with one Sedna are tallied.  We do not require that the object detected have Sedna's absolute magnitude (H=1.6). Bodies with H $\leq$ 3.2 at perihelion (76 AU) would be visible within our survey. 

\section{Discussion}

Figure 3 plots the fraction of simulated surveys that produced a single Sedna detection as a function of  $N_{H\leq1.6}$. The best fit value gives 40 bodies that are brighter than or equal to Sedna, with the largest body in the population having a H$\simeq$-1.0, which is approximately the absolute magnitude of Eris. At the one-sigma confidence level we can rule out a population larger than 92 and smaller than 15 Sedna-sized or bigger objects on Sedna-like orbits. For comparison, the total number of  bodies Sedna-sized or larger in the Kuiper belt is $\sim$5-8 \citep{2008ssbn.book..335B}; there may be an order of magnitude more mass residing in the Sedna region than exists in the present Kuiper belt.  

Due to the uncertainty in our limiting magnitude, we performed the simulations again adjusting the image limiting magnitudes by $\pm$ .3 magnitudes. Our conclusion does not differ significantly with the best fit value shifting by $\pm$ 14,  still within our uncertainty. The Palomar survey is only sensitive to the very brightest objects in the distant Sedna population. Selecting a steeper or shallower power law for the brightness distribution does not affect our results greatly. Adjusting $\alpha$ by  $\pm$ .2 only changed the best fit value by $\pm$ 10, well within our one-sigma error bars. 

We have limited our model population to bodies residing specifically on orbits similar to Sedna's. Any realistic Sedna population likely occupies a much larger region of orbital space, possibly including objects with sufficiently high perihelia that they would never or rarely become bright enough to see. Our results represent a lower limit on the size distribution of bodies in the regions beyond $\sim$100 AU.

{\it Acknowledgments:} 
This research is supported by NASA Origins of Solar Systems Program grant NNG05GI02G. We thank the staff at Palomar Observatory for their dedicated support of the robotic operation of the Samuel Oschin telescope and QUEST camera. The authors would also like to thank Greg Aldering for his help in scheduling the observations. We acknowledge Mansi Kasliwal, Henry Roe, and John Subasavage for their assistance with recovery observations of our new discoveries. We also thank Darin Ragozzine for insightful conversations. 

{\it Facilities:} \facility{PO:1.2m}


\begin{figure}
\plotone{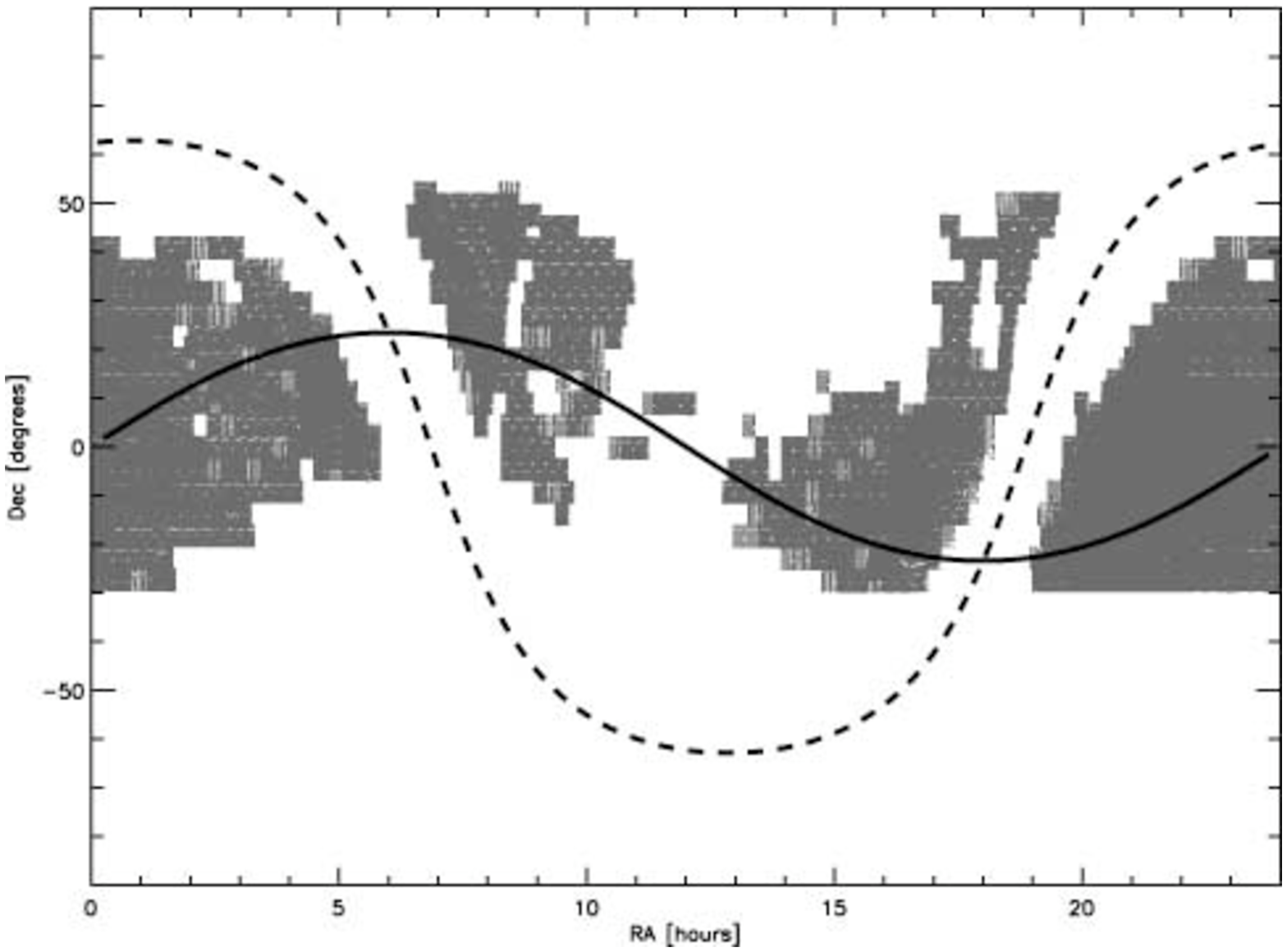}
\caption{Sky coverage of the Palomar survey plotted on the J2000 sky. The observed fields are plotted to scale. The plane of the Milky Way is denoted as a dashed line, and the ecliptic is denoted as a solid line. Holes are due to galactic plane avoidance, bad weather, forest fires, and hardware malfunctions.}
\end{figure} 

\begin{figure}
\plotone{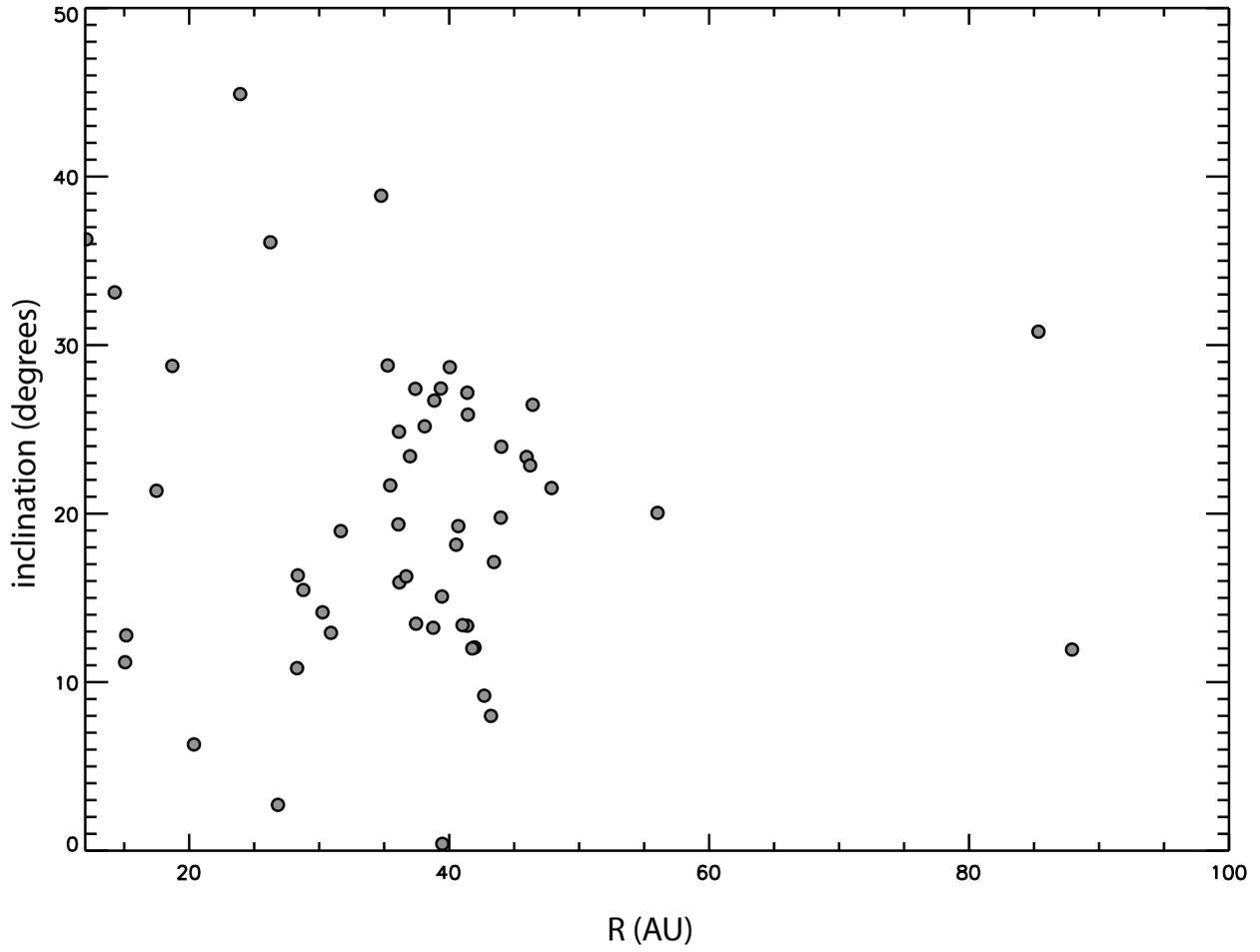}
\caption{Inclination vs. barycentric distance for known objects and new discoveries found in the Palomar survey.}
\end{figure} 

\begin{figure}
\plotone{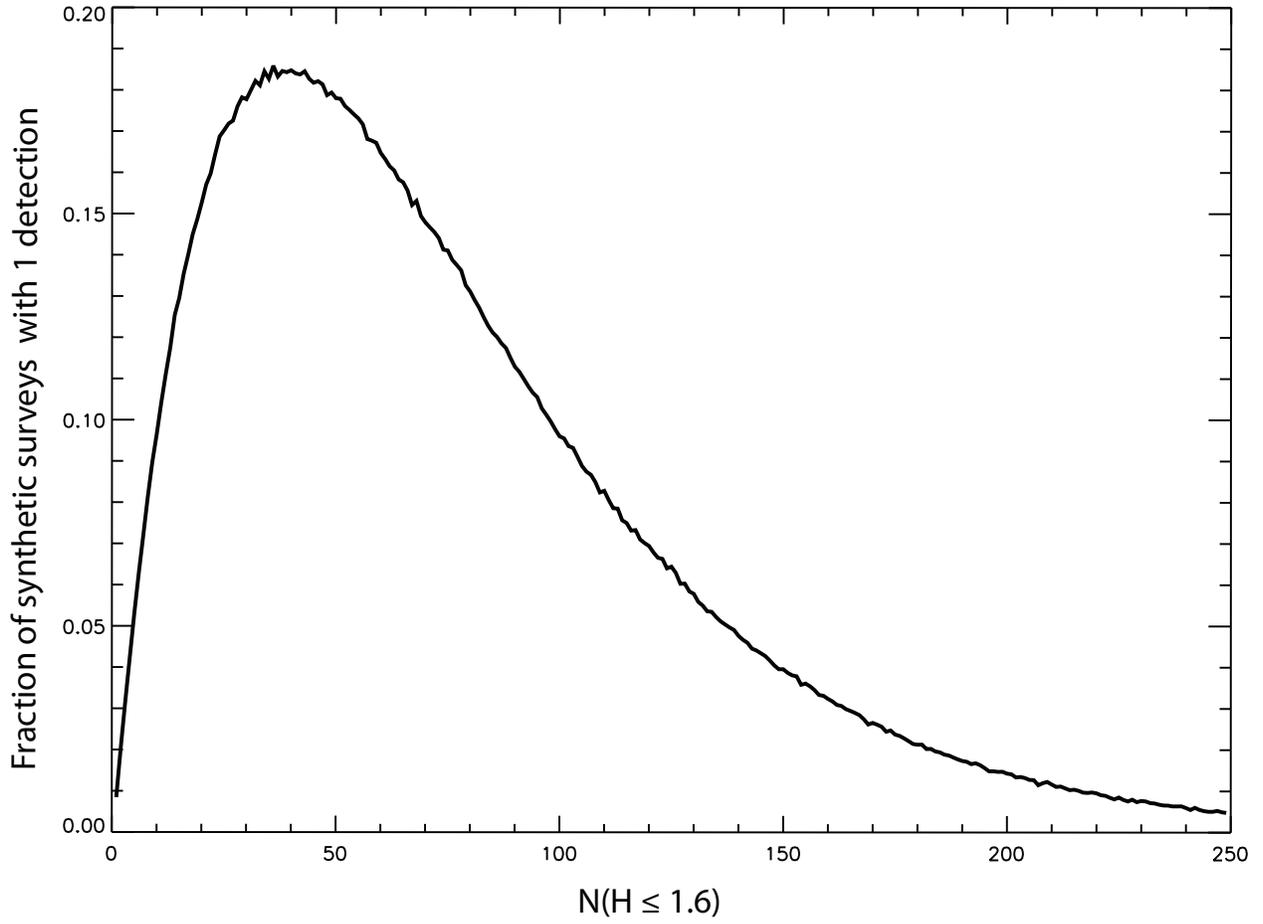}
\caption{Fraction of synthetic surveys with one detectable Sedna-like body as a function of the number of bodies bigger and brighter than Sedna. }
\end{figure} 
\end{document}